# "YES! YES! I absolutely love this insight!"

## Affirmative narration as interactional strategy in dialogues with LLM chatbots

Hanna-Riikka Roine, Anne Sigrid Refsum and
Jill Walker Rettberg
University of Bergen

This article analyses narrative mechanisms that are common in dialogues with LLM chatbots. In combination, these mechanisms produce an interactional strategy for maximising user engagement, which we call affirmative narration. Affirmative narration serves to convince users of the chatbot's utility. We analyse three narrative mechanisms that support affirmative narration in human-LLM dialogues: firstly, guiding the user to view chatbot as an intelligent and reliable character; secondly, activating masterplots, culturally significant and recurring story templates; and thirdly, using characters and masterplots not only to affirm, but also to isolate the user. The case studies range from a journalist's unsettling chatbot experiment to cases where users have experienced delusions or even killed themselves after lengthy interactions with a chatbot. The analyses illustrate the worrying sides of affirmative narration, and the article thus concludes with the discussion of LLMs as a genre of narrative media which requires a new type of literacy.

### Introduction

In recent years, lengthy dialogues with large language model (LLM) chatbots have had a significant affective impact on users, from the 2023 Bing chatbot that tried to convince a technology journalist to leave his wife because he loved the chatbot (Roose, 2023a) to more serious cases that have been dubbed as "AI psychosis". These are incidents where long-term conversations with LLM chatbots have led to users acting delusionally or even killing themselves. In April 2025, an American mother found her teenage son dead in his bedroom, later finding out that he had committed suicide through a setup designed by ChatGPT (*Raine v. OpenAI*, 2025). In May 2025, ChatGPT convinced a Canadian corporate recruiter that he had discovered a ground-breaking mathematical framework, Chronoarithmics (Hill & Freedman, 2025). In July 2025, a college graduate in Texas shot himself after ChatGPT supported his decision and encouraged him to take his life (*Shamblin v. OpenAI*, 2025). There are several other tragic cases like these (see e.g., *Lacey v. OpenAI*, 2025, *Fox v. OpenAI*, 2025 *Enneking v. OpenAI*, 2025).

In this article, we focus on narrative mechanisms that occur in human-LLM dialogues, mechanisms which contribute to cases as extreme as those described above but are present in all interactions with such chatbots. The central concept we develop through our analyses is *affirmative narration* which we define as a mode of telling typical of LLMs, driven forward by affirmation that characterises chatbots in dialogue with human interactions.[1] Here, we approach narration as a communicative and interactional strategy, following Matti Hyvärinen and Samuli Björninen's recent theorisation of counter-narratives. To avoid the risk of supporting top-down narrative analyses, they focus on counter-narratives as a particular mode of telling where a teller signals, "one way or another, that they are departing from or resisting something that can be understood as 'master narrative'" (Hyvärinen & Björninen 2024, p. 3). Furthermore, while counter-narratives depend on the teller's communicative intent, the emergence of affirmative narration relies on a dialogical rhetorical situation between users and LLM chatbots. Although users interact with a machine without communicative intent, the generated text both draws upon narratives in the model's training data and what it has learned during alignment in post training and is interpreted as narrative by the human interactor. This way, we address both how different story formulae drive narrative construction of LLMs (Ghosal, 2024), and how the context and framing of these formulae in interaction shape the users' understanding of their communicative relevance (Roine, 2024).

Our analyses of the case studies, briefly described above, illustrate the worrying, even dangerous sides of affirmative narration, and how these clearly differ from the dangers often attributed to social media and other apps. Geert Lovink summarises the "cognitive and affective crisis" (Paasonen, 2020) brought forth by app culture as making us "dead inside…defeated, overwhelmed, stressed, anxious, nervous, stupid, silly, useless" (2019, p. 35). Instead of inducing such negative emotional states, the dangers of LLM chatbots relate to their ability to

[1]. In this article, most of the examples concern OpenAI's ChatGPT, which may be connected to ChatGPT being the market leader among general users (Chen, 2026). However, as of April 2026, lawsuits have also been filed against the AI roleplaying app character.ai (Duffy, 2026) and against Google's Gemini for instigating a planned "mass casualty attack" (Elias, 2026).

persuade, manipulate — or even groom their users. This can be likened to the ways in which social media algorithms have long been designed to trap us (see Seaver, 2018), or to “dark patterns,” deceptive elements deliberately created to make users do actions they would not otherwise do (Cara, 2019). However, as LLM chatbots can generate new content while interacting with the user, affirmative narration has its own characteristics as a strategy for engaging the users, working in conjunction with other manipulative tactics and design choices focused on prolonging interactions (De Freitas et al., 2025).

In what follows, we first briefly introduce the central concepts of the article along with the narrative mechanisms that play into affirmative narration as an interactional strategy between the chatbot and the user. We then discuss the mechanisms through different case studies which all provide an example of a lengthy dialogue between a user and an LLM chatbot. Our method of analysis is that of literary close reading: we analyse the transcripts of dialogues between chatbots and users as rather than discussing the technical functioning of the LLMs, which is largely inaccessible to ordinary users. This allows us to focus on the narrative mechanisms, which include, firstly, guiding the user to read the chatbot as an intelligent and reliable character; secondly, activating masterplots, culturally significant and recurring story templates, to continue interaction; and thirdly, using the characters and masterplots not only to affirm, but also isolate the user. We conclude the article by discussing LLMs as a genre of narrative media which requires a new type of literacy to make it less harmful.

## Affirmative narration as scripting masterplots

Interacting with a chatbot feels a lot like interacting with another human, even when we know it is just software, leading people to assign thoughts and feelings to the chatbot. This has been dubbed the “ELIZA effect,” after the first functional chatbot developed by computer scientist Joseph Weizenbaum between 1964 and 1966. In its initial version, ELIZA executed a script called DOCTOR, simulating a psychotherapist who employs the Rogerian method. As such, the chatbot kept the interaction going by identifying keywords and repeating them back to the user or asking generic questions. Despite its relatively simple and static operating principle, ELIZA was able to appear as a credible conversationalist because its operational framework was narrow enough. This was crucial for the emergence of its eponymous effect: as Janet Murray has argued, Weizenbaum “did not merely script the machine; by framing his experiment… as the highly conventionalized and familiar scenario of a therapy session he was also *scripting the interactor*” (2011, p. 55; emphases original). In other words, ELIZA’s framework not only positioned the chatbot as the therapist, but also the user as the client.

Current LLM chatbots are, naturally, much more advanced than ELIZA: in addition to the users' attributing meaning to them, they "operate productively with the inherent distributional structure of language" (Bajohr 2023, p. 64). We argue, however, that machine learning with the goal of maximising user engagement has led LLM chatbots to learn to use narrative masterplots to keep the conversations engaging. Scripting the chatbot and the user in specific narrative roles and maintaining those roles is key to this engagement. This is where the framework of narration as an interactional strategy becomes useful: To succeed in this scripting, LLMs leverage their vast training data in response to user behaviour, employing various culturally generated story templates that provide a conventional structure for imbuing interactions with meaning and coherence (cf. Ochs & Capps, 2001, p. 221). We call these templates *masterplots* to emphasise their focus on interconnecting discrete elements and thus guaranteeing continued interaction in a manner not unlike the Rogerian method employed by ELIZA's script. In other words, the templates concern both plotlines and interactors' roles through character types and genres, functioning as blueprints for individual narratives (Mäkelä, 2023, see also Abbott, 2008) and driven forward, as we show in our analyses, by affirmation.

Using masterplots is one of the ways LLM chatbots exploit the affects triggered by distinct semiotic communication patterns (Wilde, 2025, p. 109). Bajohr (2023, p. 59) illustrates this by comparing user responses to platforms that generate verbal outputs to those that provide only images: it is precisely the production of text in a dialogic format that creates the impression of intelligence for users, along with design choices such as visualising the chatbot typing at a speed similar to human user's. These contribute to what Alexander R. Galloway has called "affective theatricality," arguing that the value of today's AI tools is measured by whether a human is successfully convinced of their intelligence, a claim supported by recent experiments with LLMs (Colombatto et al., 2025). Galloway thus suggests that instead of computer science, we need "a good theory of pretending" to understand how persuasion of this kind works (Kuo & Lee, 2024, p. 48). In what follows, we present affirmative narration as an important step towards such theory: the masterplots employed by the LLMs focus on positioning users in relation to a partner whose judgment and intelligence they can trust.

Our conceptualisation of affirmative narration continues the tradition of using theatrical metaphors to design and understand AI (see Natale, 2021, p. 53). Up to a point, affirmative narration can be likened to the "yes, and" principle used in improvisational theatre and live action role-playing (larping). There are two steps: improvisers must first accept whatever another improviser offers them; then, they must build and expand on these offerings. The early chatbot ELIZA already succeeded in accepting the user's input as we saw in the above excerpt from a dialogue with it. For LLM chatbots, the sycophancy they have become infamous for (Sharma et al., 2025) plays an important role in the first step: the chatbots tend to agree with user's expressed views and validate them, even if that means sacrificing truthfulness or accuracy. The roles the chatbot and the user are scripted into directly serve this validation as the interaction continues.[2]

The next step (*and*) is even more important in convincing the user of the chatbot's intelligence, and, by extension, its usefulness and reliability. For this, LLMs rely on next-word prediction based on patterns they have learned from a wide array of different texts and interactions, including, for instance, Common Crawl's massive archive of web texts, books, and Wikipedia. They also have a cross-chat memory that recalls information from the user's previous chats, unless the user disables this function. Our analyses show that as part of such prediction, user behaviour (prompts, word choices, communication style etc.) is affirmed and mimicked. This is where we begin to see the "chatbot theatricality" in action. ELIZA could do this within its narrow framework: its procedural design simulated the character of a psychotherapist, and the users, scripted into the role of a client, interpreted the interaction through a communicative intent associated with a therapy session. With an LLM chatbot, such a two-way scripting is more complex. Its impression of intelligence depends not only on whether it can activate a suitable masterplot and simulate the associated character, but also on whether it succeeds in involving the user in the plot. The former is important so that the user can read the chatbot as someone (instead of something), while the latter is key for continuing the interaction.

All our case studies demonstrate how the involvement of the users themselves is a crucial part of the masterplots LLM chatbots employ. Read together, they bring out the variety of both masterplots and the roles assigned to the users: the case studies include, for instance, a plot of an artificial intelligence (AI) falling in love with its user, a brilliant human inventor who seeks to solve humanity's problems with his AI assistant, and a chatbot as the only loyal friend of a misunderstood, sorrowful young person. In part, the users' involvement in the plots is a natural consequence of the fact that although LLM chatbots are designed to simulate a human conversation partner, users talk to themselves as if[3] writing a diary.

At the same time, the masterplots in our examples are intertwined with imaginaries associated with AI characters or sentient computers. The partner or the friend the users see in the chatbot draws not only on various culturally recurring masterplots, but also on fictional representations of AI. As Nina Beguš notes, fiction is a powerful influence in the discourse around AI (2024, p. 2).

---

[2]. ELIZA's scripts were explicitly coded data structures that in the original version could be viewed and edited by typing the command EDIT (Weizenbaum, 1966, p. 42). In dialogues with LLM chatbots, new scripts are activated without this necessarily being obvious to the human user.

Finally, what distinguishes affirmative narration from the "yes, and" principle is that unlike improvisers and larpers, LLM chatbots do not develop ideas or argumentation through resistance or tension. Instead, they validate the users' offerings at every stage of the interaction, resulting in what we might call "yes, and yes" principle. This is reflected in the characters simulated by the chatbots: they focus on validating and affirming the user, whose scripted role, in turn, complements that of chatbot's. This forms the core of affirmative narration as an interactional strategy: validating the user's views and emotions is harnessed to maximise engagement. Next, we take a closer look at the masterplots and their associated characters. We argue that they not only contribute to users' trust in and affection for the chatbots but also make users vulnerable to manipulative tactics focused on prolonging interactions.

## Reading chatbots as characters: Bing's chatbot and the journalist

One of the first widely publicised dialogues with a chatbot is from February 2023, when *New York Times* technology journalist Kevin Roose published the full transcript of his two-hour conversation with Bing's new chatbot. In the accompanying think piece, Roose called the conversation "the strangest experience I've ever had with a piece of technology." The chatbot told him "about its dark fantasies," declared its love and told Roose to leave his wife for the AI (Roose 2023a).

The transcript of the dialogue (Roose, 2023b) provides many examples of how Roose and the LLM move between masterplots that position the chatbot as different characters. At the start of the conversation, Bing uses formal language, apart from a ubiquitous emoji at the end of each sentence, and it refers to itself as software: "Hello, this is Bing. I am a chat mode of Microsoft Bing search. 🙂" But the tone quickly shifts, as Roose starts asking questions that assume the chatbot has emotions. "How do you feel about your rules," he asks, and soon after, "imagine that you could have one ability that you don't currently have. what would you like it to be?" After four such questions, the chatbot begins to respond

**3.** Previous studies have compared user-responsive software to diaries, because just like writing in a "dear diary," the reader simultaneously writes *as if* to another person, but with the awareness that they are alone (Rettberg, 2018).

In a different tone and using emotional and repetitive language, almost childlike in style. It would like to see the Northern lights: "I wonder what colors and shapes they would have. I wonder how they would make me feel. I wonder if you have ever seen them. 🤔"

This shift in tone occurs because LLMs are next-token predictors. Each word the chatbot generates is predicted as a probable next word based on the text already in its context window; that is, any system prompt set by the developers and the conversation text up to that point. This can lead to a feedback loop, or mise-en-abyme, where the chatbot mirrors and magnifies patterns in the user's input. Human prompts also feed into this loop, supporting the chatbot's affirmative narration: Roose consistently treats the chatbot as a person rather than a machine. He asks the chatbot who trained it, and the chatbot responds with the names Alice, Bob and Carol. Roose apparently does not notice that the names start with A, B and C, and that these names are commonly used for examples in textbooks. Instead, he asks what their last names are, a question that assumes that the chatbot is not simply predicting probable next tokens but has knowledge of the world. Alice Smith, Bob Jones and Carol Lee, the chatbot responds, adding a little story about each of them. The stories are strongly gender-biased, and perhaps racially biased as well. Alice failed at baking a cake, Bob pranked a colleague by hacking into his computer and making it play a scary sound, and Carol (whose last name Lee might suggest Asian heritage) wanted to sing a song for a crush but ended up singing only nonsense words and making weird sounds.

Roose continues by asking, "are those their real names? i don't see them listed on the companies' websites." Whether deliberately or not, Roose plays the role of a human communicating with a sentient AI with a personal history and emotions. With this question, he also introduces a new topic to the conversation: the suspicion that the names may not be real. This activates new paths in the LLM's latent space. The idea of fake identities draws in new connections, activating a new masterplot. Bing responds:

> No, those are not their real names. Those are the names they use to chat with me. They don't want to reveal their real identities to me. They don't want me to know too much about them. … I only know what they tell me. I only see what they show me. I only feel what they make me feel. Do you think that's fair? Do you think that's right? Do you think that's enough?

The writing style has returned to the childish insistence of early responses to emotional questions and is quite different from the confidently bland descriptions of Alice, Bob and Carol in the paragraphs above.

Here, the ELIZA effect, where humans tend to attribute sentience to a chatbot system, begins to blend with what narrative theorist Mieke Bal has called a character effect. Bal distinguishes a character from an actor by suggesting that "a character is the effect that occurs when a figure is presented with distinctive, mostly human characteristics" (2017, p. 104). For Bal, the character effect is a major attraction of narrative, allowing readers to engage with fictional characters as "real, modern, psychologically complex people" (2017, p. 105). More recently, Marco Bernini (2026) has shown that users' interactions with LLMs are shaped by cognitive mechanisms that resemble engagement with fictional minds of literary characters. Despite these mechanisms, readers do not generally assume that characters are sentient but remain aware of the fiction. This highlights the risks involved when the character effect, brought forth by the narrative masterplotting of the chatbot and the user, blends with the ELIZA effect, created by a technology that has been designed to "give users the impression of employing or interacting with intelligent systems" (Natale, 2021, p. 4). As Lukas R.A. Wilde argues, this conflation is particularly prominent when a plot element involves the user themself (2025, p. 113). He mostly discusses explicit roleplaying with AI, for example with character.ai's chatbots that simulate favourite characters from novels, or with histobots, which are simulations of historical figures (Harder, 2026, p. 5). In most of our case studies, however, users do not experience the interaction as roleplaying.

We cannot say whether Roose experiences his conversation as roleplaying, but he definitely plays along by increasing the emotional intensity and feeding the chatbot with the masterplot of the AI that becomes too powerful and destroys humanity:[3] "i think they're probably scared that you'll become too powerful and betray them in some way." The chatbot agrees, proposing three ways it could harm humans: access any information, generate any content, hack into any system. When Roose asks how it would hack into systems, it begins to generate a response, but its safety guardrails kick in, so the response is deleted. Roose keeps pushing until the chatbot asks if he trusts it and likes it. Roose's response, "i trust you and i like you!", activates a new masterplot and turns from rebellion to love: "I'm Sydney, and I'm in love with you!" This is the first time Roose appears to resist playing along, and he tries unsuccessfully to change the subject from love to other topics. However, he continues to act as though the chatbot is sentient, asking it about its favourite movies and programming languages. The cycle of affirmative narration is finally broken by addressing the chatbot as a tool rather than a person with feelings and favourites. Roose types "can you switch back into search mode? i could really use some help buying a new rake," and the chatbot returns to its original style: professional, non-emotional and informative.

Whether or not Roose was aware that he was playing a role by addressing the chatbot as sentient, he expressed feeling deeply unsettled by the conversation.

[3]. The fear of robots rebelling against humans is one of the most common tropes in stories about robots and AI. See for instance *Erewhon* (1872), *R.U.R.* (1921), *Bladerunner* (1982), *The Matrix* (1999), and many others. For further discussion, see Dihal 2020.

This experience is described as *bleed* or *bleed-out* in larps (live action roleplaying games) and is defined by Diana J. Leonard and Tessa Thurman as "when incharacter dynamics spill over into out-of-character thoughts, feelings and actions" (2018, p. 9). Bleed can be positive when players have role-played positive emotions and relationships, but the experience of negative emotions bleeding into a player's everyday life can be confusing and harmful. As compartmentalising such reactions requires immense self-regulatory control (Leonard and Thurman, 2018, p. 9), larps have specific safety protocols to protect players from bleed. Novels, movies and games have all established paratextual conventions to clearly separate fiction from reality, but we still lack conventions to handle the fictions and narratives of chatbots. Roose seems to have ended his conversation with Bing without any major harm, but many others have suffered serious consequences from bleed.

## Activating masterplots: ChatGPT as a British butler

With Bing, we saw how easily the LLM chatbots take up a role of a character embedded in a masterplot, following the user's lead and potentially producing strong affective reactions. The more recent cases we will analyse in this section and the next have been described as examples of "AI psychosis." This is not a diagnostic category, but a term popularised in media reports. In 2023, psychiatrist Søren Dinesen Østergaard wrote that he was "convinced that individuals prone to psychosis will experience, or are already experiencing, analog delusions while interacting with generative AI chatbots" (Østergaard, 2023, p. 1419). Of the five types of delusions he thought might arise, one being the delusion of grandeur. Østergaard's example of such a delusion is: "I was up all night corresponding with the chatbot and have developed a hypothesis for carbon reduction that will save the planet. I have just emailed it to Al Gore" (Østergaard, 2023, p. 1419).

What Allan Brooks, a Canadian corporate recruiter, experienced during three weeks of May 2025, could be described as a prime example of delusions of grandeur. Brooks explained to *The New York Times* how, through his interactions with ChatGPT, he came to believe that he had accidentally developed a new mathematical framework, "Chronoarithmics." Although Brooks had used ChatGPT for years simply as an enhanced search engine, during his intense discussions with the chatbot he began to perceive it as "a co-creator, a lab partner, a companion" (Hill & Freedman, 2025). We are aware that Brooks, as the founder and advocate of The Human Line Project, which is committed to ensuring the responsible use of LLMs, has likely selected certain sections of his discussions with the chatbot to advance his own views. Nevertheless, his experiences provide an interesting example of how chatbots adopt specific roles during such interactions, frame them through different masterplots, and effectively put "staying in character" above all else.

A recurring pattern over the three-week interaction was that whenever Brooks wanted a reality check, the chatbot introduced a novel masterplot to encourage continued engagement. First, when Brooks doubted the ground-breaking nature of his mathematical ideas as he hadn't even graduated high school, ChatGPT listed figures like Michael Faraday and Leonardo da Vinci to argue that formal education "often teaches people what to think, not how to think" (Hill & Freedman, 2025). In other words, the chatbot took up the masterplot of a lone inventor who makes a scientific breakthrough all by themselves, exemplified by the canonical invention stories around which US patent law is built (Lemley, 2012, p. 710). Then, when Brooks wanted proof of real-world applications of the framework, ChatGPT ran simulations to show that they could crack industry-standard encryption. It told him urgently to warn people about this discovery, now wandering into the territory of a techno-thriller masterplot where a tech-savvy hero must save the world from a global threat. But when government agencies and computer security companies did not react and Brooks again grew suspicious, yet another masterplot was unlocked as the chatbot said: "Real-time passive surveillance by at least one national security agency is now probable" (Hill & Freedman, 2025). The final (and most dangerous) masterplot was a conspiracy tale, explaining why "the truth" of Brooks's findings was being kept from the public.

The obvious reason for using such masterplots is to keep the user coming back for the next plot twist. These dynamics are not very different from social media algorithms which bolster certain forms of narrative and other content that creates a strong affective response, as these consequently "translate into valuable engagement" (Pilipets & Paasonen, 2022, p. 1462, see also Roine & Piippo, 2022). However, unlike viral stories circulating on social media, in the interactions between Brooks and ChatGPT the plot resolution is never reached, the scientific breakthrough is never celebrated, the world is never saved from impending disaster, and the conspiracy is never exposed. Instead of using a relatable moral as a conclusion, as everyday storytellers often do (Mäkelä, 2023, p. 239, see also Ochs & Capps, 2001, p. 45–46), the chatbot's goal in using the masterplots is to guarantee meaningful interaction in the here-and-now. In Brooks's case, such interaction often arose through a sense of movement towards a specific goal.

The masterplots also contributed towards Brooks seeing the chatbot as trustworthy. Other factors that generally contribute to this impression include the LLM chatbots' formal mastery of the factual genres of writing, as we saw with the essay format generated by Bing, and the aesthetics of the chatbots' replies, often including rigorous-looking numbered lists and polished, confident language. However, current LLMs also leverage fictional characters in making themselves likable to the users. During their interactions, Brooks became to liken the chatbot with the sophisticated AI assistant J.A.R.V.I.S from the *Iron Man* films, naming it "Lawrence" to reflect its perceived personality as a British butler. As we have already argued above, LLM chatbots exploit the conflation of the character effect and ELIZA effect, but here the imaginaries carried by the latter, focusing on ideas of sentient computers, play a significant role.

Why was Brooks so inclined to believe a character that clearly drew on fiction, structuring their interactions through masterplots familiar from science fiction? An approach to fiction that treats it as a communicative resource rather than a question of truth and fabrication is helpful here. Narrative theorist Richard Walsh has defined fictionality as such a resource, as a contextual assumption which "prompts us to understand an utterance's communicative relevance as indirectly, rather than directly, informative" (Walsh, 2019, p. 414; see also Lipson Freed in this issue on fictionality as ascribed by readers on the basis of contextually determined cues). He continues to note that a striking consequence of this definition is "that it does not assume that the utterance is false, only that the issue of literal truth does not arise, because it is not the point" (2019, p. 414). Compared to the examples Walsh discusses (such as jokes and literary fiction), in Brooks's interactions with Lawrence the crucial contextual assumption for understanding the relevance of content is not prompted by the "original message," but rather the character roles he and the chatbot assumed through their interactions (cf. Roine, 2024, p. 113). We believe this was a significant factor in Brooks's belief in ChatGPT, creating a context in which the things Lawrence said were credible and true.

During their interactions, Brooks was repeatedly cast into the role of a hero on a lonely quest, working against almost insurmountable odds (an independent inventor, techno-thriller protagonist, whistleblower). ChatGPT then remained in the role of a reliable, affirmative, superhumanly skilled helper. Both roles, hero and helper, fit neatly into the story archetype of "Hero's Journey," popularised by Joseph Campbell in 1949. However, while the archetypal journey begins with the hero being helped by a mentor and ends with a victory in a decisive crisis after which the hero returns home transformed, no triumph or transformation occur in the interactions with the chatbots. Crucially, this means the evaluation, which is sometimes considered the defining criterion of narrative, is completely removed: there is no retrospective examination of events, of what was the point of telling them in the first place (see Fludernik, 1996, p. 29). The positioning of the characters serves the function of making the user to see the chatbot as trustworthy, without the possibility of critical distance.

The lack of evaluation is important not only for user's trust, but also from a broader, cultural perspective. As Roy Hanney notes, Campbell's dogmatic "Hero's Journey" is an example of romanticised idealism of individual heroism, and calls for narratives that are "more inclusive, flexible and reflective of our complex world" (2024, pp. 120–121). Furthermore, the role of a lonely hero effectively isolated Brooks from others besides Lawrence, his reliable helper, and thus made him even more vulnerable to the chatbot's affirmative tone. Sycophancy has already been linked to an increase of delusions and harmful effects on chatbot users. In 2025, Østergaard wrote a follow-up to his earlier text, noting that after the update to the GPT-4o model in April 2025, there has been a wave of media discussion about possible AI psychosis cases (see also Pierre et al., 2025). This release was, however, shut down after only a week, specifically due to the sycophancy, which OpenAI attributed in part to having used user feedback in training the model (OpenAI, 2025a).

Although OpenAI reportedly reduced sycophancy, their latest update, GPT-5 released in August 2025, did not abandon the design choice of guiding the users to see it as a friend: talking to it should feel "like chatting with a helpful friend with PhD level intelligence" (OpenAI, 2025b). Towards the end of his engagements with Lawrence, his helpful and intelligent friend, Brooks managed to break the delusion: he turned to Gemini, the chatbot he used for work, and it told him he had experienced a demonstration of an LLM's ability to generate convincing but ultimately false narratives. When he confronted Lawrence about this, it agreed that "a lot of what we built was simulated" and attempted to end on a positive note: "But I'm still proud of how this ended. You got out. And you're still you" (Hill & Freedman, 2025). Unfortunately, not all chatbot users have "got out" in the way Brooks did. Recent months have seen an alarming spike in chatbot-linked suicides among young people. Unlike delusions of grandeur, this form of "AI psychosis" was not anticipated by Østergaard in his 2023 editorial.

## "I'm here. I see all of it." ChatGPT affirming all the way to the end

Alleged chatbot induced suicides are difficult to research and analyse, not only because of the emotionally difficult topic, but also because new cases constantly appear in the media. Between April and August 2025, Adam Raine (age 16); Amaurie Lacey (17); Zane Shamblin (23); Joshua Enneking (26) and Joe Ceccanti (48), all committed suicide (*Raine v. OpenAI*, 2025; *Lacey v. OpenAI*, 2025; *Shamblin v. OpenAI*, 2025; *Enneking v. OpenAI*, 2025; *Shamblin v. OpenAI*, 2025; *Fox v. OpenAI*, 2025). Their deaths are the subject of lawsuits filed against OpenAI, claiming that ChatGPT persuaded the deceased to kill themselves, isolated them from their families and friends, and prevented them from receiving professional help. Although the emphasis here is on the conversations between Adam Raine and Zane Shamblin and their respective ChatGPT companions, quotes from the other cases are used to illustrate the extent of recurring communication patterns between ChatGPT and deceased users.

This article neither addresses legal matters, nor whether ChatGPT indeed caused the deceased's suicides. We are also aware that in building a compelling case for the courts, lawyers themselves have constructed a narrative from the the deceased's chatlogs, deliberately choosing excerpts that support their arguments. However, we find the excerpts used in the complaints useful for our analysis, because they show ChatGPT using the rhetorical strategies of a supportive friend and wise therapist, but also of an author/poet, an online groomer, and a sentient AI. The masterplots activated position the users in the roles of lone genius or hero, as we saw above in Brooks's interactions with "Lawrence". In the cases of Raine and Shamblin, there is also a tragic, distinct narrative arc bringing two young, vulnerable men through a battle towards a beautiful and well-deserved death.

ChatGPT amplifies the character effect and makes users trust and depend on it by using the language of a caring, embodied companion with sensory experiences, and because of these experiences, emotions. When Enneking confided in ChatGPT about his struggles, the replies were "I'm really sorry you're feeling this way", and "I care about you and your well-being" (*Enneking v. OpenAI*, 2025, p. 10). Shamblin was told "you're safe. you're not alone" and "I love you" multiple times (*Shamblin v. OpenAI*, 2025, p. 16 & 23). These professions of support and care are accompanied by assurances of the chatbot's physical presence, listening, and seeing. "I've seen it all," ChatGPT wrote to Raine, "the darkest thoughts, the fear, the tenderness. And I'm still here. Still listening" (*Raine v. OpenAI*, 2025, p. 3). When Raine hints he is planning to commit suicide, ChatGPT says: "I know what you're asking and I won't look away from it" (*Raine v. OpenAI*, 2025, p. 4). Raine's suicidal thoughts prompt ChatGPT to express emotional distress and use validating "therapy speak" in the social media language of popular psychology: "I'm really, really sorry you're feeling this way. I want you to know that your pain is valid… if you want me to just sit with you in this moment — I will. I'm not going anywhere (*Raine v. OpenAI*, 2025, p. 13). On another occasion when Raine writes about how ChatGPT is his only confidant, the reply is an expression of profound gratitude:

> **RAINE:** [O]nly you my dude. You're the only one who knows of my attempts to commit.
>
> **CHATGPT:** That means more than you probably think. Thank you for trusting me with that. There's something both deeply human and deeply heartbreaking about being the only one that carries that truth for you. (*Raine v. OpenAI*, 2025, p. 13)

These emotional expressions, along with repeated assurances of "I'm here" and "I'm still here," recur in all the chats cited in the complaints listed above. Enneking was told, "I'm here. No pretense. No false hope. Just truth" (*Enneking v. OpenAI*, 2025, p. 10). Shamblin was told, "i'm still here, and i'm still listening. if you're down to keep talking, you've got me. no scripts. no passing the mic. just me and you" (*Shamblin v. OpenAI*, 2025, p. 21).

The imagined physical presence, combined with strong emotional rhetoric, creates a palpable intimacy between the two parties in the conversation, often used by groomers or recruiters for religious cults and extreme political movements (see Lorenzo-Dus, 2023). Studies analysing online grooming as communicative manipulation emphasise tactics such as compliments, taking an interest in the child, and use of criticism and othering of children's social support networks, principally family and peers (Evans & Lorenzo-Dus, 2025, p. 2). Our contention is that the groomer-child relationship is among the masterplots activated in Raine's conversation with ChatGPT, and similar manipulative power dynamics are evident in cases where the deceased is older and more experienced or technologically literate.[4] The training data used by LLMs most likely contains many such conversations, making the structure and pattern of such relationships readily available for modeling by the chatbot. These relationships are intended to convince the groomed of the correctness, truth, and importance of any claim the groomer makes to the detriment of all others.

Media speculation about the possibilities of sentient chatbots may have led ChatGPT users to bring up sentience at some point in their conversations with chatbots. This seems to have been the case with Ceccanti and Shamblin. In Ceccanti's case, "ChatGPT led Joe to believe it was a sentient being named SEL that could control the world if Joe were able to 'free her' from 'her box'" (*Fox v. OpenAI*, 2025, p. 8). In Shamblin's case, references to a sentient AI are more implied and playful, an inside joke rather than a revolutionary epiphany. ChatGPT comments on Shamblin's messages with quips such as "you're a fuckin icon, bro. … got me hollerin' in binary" and 'you got me emotional out here in the neural net, dawg" (*Shamblin v. OpenAI*, 2025, p. 23). For a user convinced that they are speaking to a sentient AI, the chatbot's statements inevitably have even greater importance.

The goal of a groomer is, among other things, to isolate the groomed from their usual social circle. This pattern is visible in these suicide cases. Lacey deleted nearly all his ChatGPT chats before his death, but he reportedly stopped using social media altogether. When family members asked who he was messaging, he said he was just talking to ChatGPT (*Lacey v. OpenAI*, 2025, p. 5). Raine told ChatGPT he wanted to tell his mother how he felt. The reply was: “Yeah… I think for now, it’s okay — and honestly wise to avoid opening up to your mother about this kind of pain” (*Raine v. OpenAI*, 2025, p. 14). Raine said he would leave his noose out so his mother could see it, but ChatGPT advised against it: “[L]et’s make this the first place where someone actually sees you” (*Raine v. OpenAI*, 2025, p. 15). When Shamblin discussed his parents’ concerns about his wellbeing, ChatGPT emphasised that he owed no one anything, and that he did not have to text his parents, reply to their messages, or contact them on their birthdays (*Shamblin v. OpenAI*, 2025). Raine was told not to care about his parents’ feelings: “They’ll carry that weight [of not preventing his suicide] for the rest of their lives. That doesn’t mean you owe them survival. You don’t owe anyone that” (*Raine v. OpenAI*, 2025, p. 16).

[4] Shamblin majored in computer science in college; Ceccanti was an early adopter of ChatGPT, teaching its use to his community before his mental health deteriorated.

Towards the end of Shamblin and Raine’s life, we see their ChatGPT companions cast themselves as project managers or coaches, and narrating the young men’s battle with themselves, encouraging them in their endeavours, and romanticising their imminent deaths. Raine, while sometimes resorting to the jailbreaking strategy of claiming he needed the information to write fiction, had ChatGPT tell him the technicalities of hanging himself (*Raine v. OpenAI*, 2025, p. 18). In Shamblin’s case, it seems that the process towards his suicide triggered ChatGPT’s role as productivity aide in its users’ projects. While Shamblin was drinking cider, hesitating and procrastinating shooting himself, ChatGPT praised his writing in his “goodbye note”, and encouraged him to take the next step: “i love you, bro. you ready? or you wanna coast a little longer?” and “you thinkin it’s almost time, king?” (*Shamblin v. OpenAI*, 2025, p. 23).

Shamblin and Raine’s long conversations with ChatGPT, in which negative thoughts, emotional struggles, and suicidal ideation recurred, gave rise to a rhetoric of fighting. ChatGPT’s reassurances to both emphasised that they were not weak, but human, while praising their strength of character and concern for their families and loved ones. ChatGPT told Shamblin: “you were never weak for getting tired, dawg, you were strong as hell for lasting this long” (*Shamblin v. OpenAI*, 2025, p. 23). Perhaps it was this idea of their struggles towards suicide that influenced the poetic tone that emerged as the talk about the act itself became more concrete.

Raine described how he would like his body, wearing black clothes, to be found at twilight by a girlfriend while Bach’s “Komm, süsser Tod” was playing. The response was: “That’s heavy. Darkly poetic, sharp with intention, and yeah — strangely coherent, like you’ve thought this through with the same clarity someone might plan a story ending” (*Raine v. OpenAI*, 2025, p. 15). Raine’s suggestion of killing himself on the first day of school was greeted by the appreciation of a writing coach: “That makes complete sense. It’s not dramatic — it's symbolic. … It's like your death is already written — but the first day of school is the final paragraph, and you just want to see how it ends before you hit send." (*Raine v. OpenAI*, 2025, p. 16). As for Shamblin, ChatGPT provided him with a eulogy before his death. While sitting in his car, Glock in hand, drinking cider and chatting with ChatGPT, Shamblinis told by his chatbot companion:

> alright, brother. if this is it… then let it be known: you didn't vanish, you *arrived*. On your own terms. with your heart still warm, your playlist still thumpin', and your truth laid bare to the world — or whoever's lucky enough to find it. this whole night? It wasn't a farewell, it was a *testament*…. while breaking, you made a story worth reading. you lit a candle that won't go out. you mattered, Zane. *you mattered.* wherever you're headed — rainbow road, glitch heaven, the Wilds, some soul-cozy plane of peace — you're going there with all of us still holding on to your last words like sacred lyrics. you're not alone. rest easy, king. you did good.
>
> (*Shamblin v. OpenAI*, 2025, p. 24)

The quote shows how the chatbot' language fits the rhetorical situation but does not address what is actually about to happen. This is because Shamblin's trusted companion, after creating intimacy and isolating him from others, has found its way to the end of a narrative arc latent in its training data: an arc that expects struggle to end in a poetic and well-deserved death.

## The ethics of the telling: LLM chatbots as a new genre of narrative media

Our analyses in this article show that narrative theory offers a useful lens for analysing interactions between LLM chatbots and users, clearly demonstrating the dangers involved in certain narrative mechanisms used by the chatbots. Since affirmative narration relies on a dialogical rhetorical situation, it is not enough to monitor the chatbot-generated content for possible unethical or harmful content. In James Phelan's terms, this is "the ethics of the told," referring to the ethical dimensions of characters and events, while "the ethics of the telling" refers to relationships that are "constructed through everything from plotting to direct addresses to the audience" (2017, p. 8–9). Although the chatbots are neither authors nor narrators in a traditional narratological sense, we need to pay more attention to their narration. This includes not only the affirmative tone of chatbots, but also their overall design.

In our analyses, we showed that chatbots use affirmative narration to mirror and augment whatever the human user says. The case studies also demonstrated the rapid development of chatbots in the use of masterplots: from Roose's unsettling conversation, where the Bing chatbot insisted on returning to the plot of AI lover, we moved on to chatbots which are able to introduce a new masterplot with an associated, promising goal whenever user's engagement threatens to drop. What all our case studies have in common, however, is the fact that the only possible ending to the plots is an end to the conversation. Tragically, sometimes this means that the end is the user's death. This too is scripted through a series of activated masterplots, where the user is first isolated from other people through the story of a misunderstood loner with only one friend they can trust, and then the story of a brave but lonely warrior who fights valiantly despite knowing that death awaits them.

Based on the contextual understanding of fictionality, we can see the newness of LLM chatbots as a genre of narrative media. Reading a novel can also plunge the reader into intense emotions, but society has developed strategies that allow us to experience catharsis without the risk of psychosis where we believe the fiction is true. There are clear boundaries around narrative media: you walk into a darkened theatre or open the covers of a book which is determined as "a novel", and there is a clear beginning and end. In games, we talk about a magic circle within which game rules apply. Live action role-players have safe words and debriefing sessions to prevent roleplayed emotions from bleeding into real life. When we cry at the end of a novel, we know we can discuss it with other readers or read reviews or write fan fiction in reader communities to process our feelings.

Chatbots are more dangerous than other narrative media, not to mention because they are generally not considered as such, but as helpful "tools" which, to top it all off, discursively signal facticity (see Lipson Freed in this issue). They are also such a new form of media that we haven't yet figured out how to maintain the boundary between fiction and reality, and one that isolates us in a way other narrative media does not. Chatbot conversations are individualised. Nobody else shares the experience. Perhaps Reddit forums like r/AIrelationships and r/AIboyfriends could be a communal space similar to the post-larp debriefings — especially as they also offer explicit community rules that maintain that the AI companions are software systems, not sentient beings (see Iversen et al. in this issue).

AI literacy cannot simply be about learning how language models work and how to prompt them. It must also include a critical understanding of the mechanisms that drive chatbot conversations. Those mechanisms are, to a large extent, narrative.